# Theory of Tunneling for Rough Junctions

M.B. Walker

*Department of Physics, University of Toronto, Toronto, Ont. M5S 1A7*

(October 16, 2018)

A formally exact expression for the tunneling current, for its separation into specular and diffuse components, and for its directionality, is given for a thick tunnel junction with rough interfaces in terms of the properties of appropriately defined scattering amplitudes. An approximate evaluation yields the relative magnitudes of the specular and diffuse components, and the angular dependence of the diffuse component, in terms of certain statistical properties of the junction interfaces.

PACS numbers: 73.40.Gk, 74.50.+r

INTRODUCTION The study of the quantum mechanical tunneling of electrons between two metallic electrodes separated by a thin barrier is an important method for investigating condensed matter systems (e.g. see Ref. [1]). Although the vast majority of tunneling experiments have been carried out on tunnel junctions whose interfaces have a significant roughness, the impressive theoretical literature [1] treating the properties of different types of tunnel barriers and tunneling mechanisms has almost without exception (see however Ref. [2]) discussed only the case of flat tunnel junctions. This article presents the first detailed theory of tunneling appropriate for tunnel junctions with rough interfaces. The potential significance of such a development is apparent from one of our conclusions, namely that for junctions where the interface roughness fluctuations exceed an electron wavelength in magnitude, the contribution of the diffuse transmission of electrons to the tunneling current dominates the specular transmission that is usually calculated.

A central idea in the flat interface theory of tunneling is that for thick barriers the electrons which dominate the tunneling are those whose momenta are directed close to the forward direction [1,3–5]. This "tunneling cone" effect is the basis for attempts to determine the anisotropy of the superconducting energy gap (see page 126 of Ref. [1]), and has also recently been invoked in the explanation of tunneling phenomena high temperature superconductors [6–8] where the spectrum of quasiparticle excitation energies is highly anisotropic. The investigation carried out below of tunneling directionality in the case of rough interfaces (where flat interface tunneling cone ideas are not applicable) thus has important implications for these studies.

The theory of wave scattering at rough surfaces is a highly developed subject [9] with applications in many areas of physics. Below, some established ideas from these studies, such as the use of certain scattering amplitudes and of ensemble averages over the random variables describing the rough interfaces, are used to derive a formal expression for the tunneling current and to separate it into specular and diffuse components. This expression is then evaluated within the framework of two complementary classical approximation schemes, a small perturbation method valid for roughness fluctuations smaller than the electron wavelength, and a quasiclassical approximation (implemented via the tangent plane method) valid in the opposite limit. The approach of this article is thus quite different from a previous discussion of diffuse scattering in tunneling [2] which has no way to separate the specular from the diffuse scattering, to calculate their relative magnitudes, or to investigate the factors influencing directionality in the case of rough interfaces.

The SUMMARY AND CONCLUSIONS section at the end of the paper gives an overview of the main results.

FLAT TUNNEL JUNCTION INTERFACES AND THE TUNNELING CONE Consider an electron tunneling from one metal to another through an insulating barrier. In the prototypical problem [1,4] the electron is described by the Schrödinger equation

$$(-\hbar^2/2m)\nabla^2\psi + V(z)\psi = E\psi. \qquad (1)$$

The potential $V(z)$ is shown in Fig. 1. The energy of the electron $E$ lies between 0 and $V_0$ so that the insulating slab is a classically forbidden region.

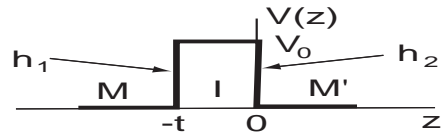

FIG. 1. Barrier potential for an MIM′ (Metal Insulator M′etal) tunnel junction. The random height functions associated with interfaces at $z = -t$ and $z = 0$ are $h_1(\mathbf{r})$ and $h_2(\mathbf{r})$, respectively.

The electron wave function in the insulating region has the form

$$\psi_I(\mathbf{R}) = e^{i\mathbf{k}\cdot\mathbf{r}}(Be^{-\kappa_{Ik}z} + B'e^{\kappa_{Ik}z}), \qquad (2)$$

$$\kappa_{Ik} = [\kappa_I^2 + k^2]^{1/2}, \quad \kappa_I = [(2m/\hbar^2)(V_0 - E)]^{1/2}. \qquad (3)$$

Throughout this article three-dimensional vectors are denoted by boldface uppercase letters and two-dimensional vectors by boldface lowercase letters. Thus, $\mathbf{R} = (\mathbf{r}, z)$.

The current tunneling across a thick ($\kappa_I t \gg 1$) barrier in the presence of an applied voltage $V$ is [1]

$$J_z = \frac{2e}{h} \int dE[f(E) - f(E - eV)] \int \frac{d^2k}{(2\pi)^2} D(E, \mathbf{k}), \quad (4)$$

$$D(E, \mathbf{k}) = g(\mathbf{k})exp(-2\kappa_I t)2\pi(\Delta k_I)^2 P_{\Delta k_I}(\mathbf{k}), \quad (5)$$

$$P_{\Delta k_I}(\mathbf{k}) = [2\pi(\Delta k_I)^2]^{-1} exp\{-k^2/[2(\Delta k_I)^2]\}, \quad (6)$$

where $D$ is the transmission coefficient and $P_{\Delta k_I}$ is a normalized (i.e. $\int P_{\Delta k_I}(\mathbf{k}) d^2k = 1$) two dimensional Gaussian function of width $\Delta k_I = [\kappa_I/(2t)]^{1/2}$. The factor $exp(-2\kappa_I t)$ in Eq. 5 comes from the exponential decay of the wave function in the insulating layer, whereas the Gaussian function reflects the fact that for thick barriers, only electrons which have momenta close to the forward direction contribute to the tunneling current. (This last property is often called the "tunneling cone" effect.) The prefactor $g$ is a relatively weakly varying function of momentum and energy of order of magnitude unity, and can usually be neglected in calculations of the tunneling current (e.g. see the discussion on p. 21 of Ref. [1]). Below, $g$ and its analogues will be put equal to unity.

*ROUGH TUNNEL JUNCTION INTERFACES - FORMAL THEORY* Consider an electron which is incident on the tunnel junction from the metal M in Fig. 1. The electron passes into the classically forbidden region at $z = -t$ and then, after having been attenuated by a factor $exp(-\kappa_{Ik} t)$, it arrives at the $z = 0$ interface where it is finally transmitted into the metal $M'$. Because $\kappa_{Ik}$ is given by Eq. 3, independently of whether the interfaces are rough or not, it is primarily electrons with small parallel momentum components $\mathbf{k}$ that arrive at the interface $z = 0$. If the interface at $z = 0$ is rough, however, it will impart a random parallel component of momentum to electrons entering the metal $M'$. Thus the tunneling electrons will sample states in $M'$ having a wide distribution of parallel momentum components, in spite of the tunneling cone effect in the insulator. The essential problem is to find how the transmitted current from electrons of wave vector $\mathbf{k}$ in the metal M is distributed over the various wave vectors $\mathbf{k}'$ in metal $M'$.

The calculations below use the methods described in Ref. [9], adapted here to the problem of tunneling. Consider first a general description of the transmission and reflection of a plane wave incident from medium 1 onto a rough interface separating media 1 and 2. The wave reflected back into medium 1 is written as a linear combination of waves with all possible parallel momentum components, as is the wave transmitted into the medium 2. The wave functions in media 1 and 2 are thus

$$\psi_1(\mathbf{R}) = \frac{e^{i\mathbf{K_1}\cdot\mathbf{R}}}{q_{1k}^{1/2}} + \int \frac{d^2k'}{(2\pi)^2} S_{11}(\mathbf{k}',\mathbf{k})\frac{e^{i\mathbf{K'_1}\cdot\mathbf{R}}}{q_{1k'}^{1/2}}, \quad (7)$$

$$\psi_2(\mathbf{R}) = \int \frac{d^2k'}{(2\pi)^2} S_{21}(\mathbf{k}',\mathbf{k})\frac{e^{i\mathbf{K'_2}\cdot\mathbf{R}}}{q_{2k'}^{1/2}}. \quad (8)$$

Here $\mathbf{K}_\alpha = (\mathbf{k}, q_{\alpha k})$, $\alpha = 1, 2$ where $q_{\alpha k} = i\kappa_{Ik}$ if $\alpha$ refers to the insulating region, and $q_{\alpha k} = (K_M^2 - k^2)^{1/2}$ with $K_M = [(2m/\hbar^2)E]^{1/2}$ if $\alpha$ refers to a metallic region. The division by $q^{1/2}$ in Eqs. 7 and 8 represents the conventional normalization of the plane waves [9]. The quantities $S_{11}$ and $S_{12}$ are called scattering amplitudes.

For the tunneling problem, the basic scattering amplitudes are $S_{I,M}$ (i.e. $1 \equiv M$ and $2 \equiv I$ in Eq. 8) and $S_{M',I}$ describing the transmission of an electron from the metal $M$ to the insulator $I$, and from the insulator $I$ to the metal $M'$, respectively. An important simplification in the calculation of $S_{I,M}$ is that for the thick junctions considered here, only the exponentially decaying waves need be considered in the insulating region, and the exponentially increasing waves can be neglected (e.g. see Ref. [10]). It can be shown that in this approximation the scattering amplitude $S_{M',M}$ describing the transmission from metal $M$ to metal $M'$ is given by

$$S_{M',M}(\mathbf{k}',\mathbf{k}) = \int d^2k'' S_{M',I}(\mathbf{k}',\mathbf{k}'') S_{I,M}(\mathbf{k}'',\mathbf{k}) e^{i(q_{Ik''} - q_{Mk})t}. \quad (9)$$

The metal-insulator interfaces are given in terms of the random functions $h_\alpha(\mathbf{r})$, $\alpha = 1, 2$ by the equations $z = -t + h_1(\mathbf{r})$ and $z = h_2(\mathbf{r})$. The ensemble average of each $h_\alpha(\mathbf{r})$ is taken to be zero, so that the average interfaces are flat. The boundary conditions satisfied by the wave function are that both the wave function and its normal derivative are continuous on both interfaces.

The ensemble average of Eq. 8 must give an average wave function $\overline{\psi_2}$ corresponding to flat interfaces; the average scattering amplitude thus has the form $\overline{S}_{21}(\mathbf{k}',\mathbf{k}) = \overline{V}_{21}(\mathbf{k})\delta(\mathbf{k}' - \mathbf{k})$. The scattering amplitude is now written as the sum of its average value and a fluctuating part:

$$S_{21}(\mathbf{k}',\mathbf{k}) = \overline{V}_{21}(\mathbf{k})\delta(\mathbf{k}' - \mathbf{k}) + \Delta S_{21}(\mathbf{k}',\mathbf{k}). \quad (10)$$

Furthermore the correlation function of the scattering amplitude fluctuations can be written in the form

$$\overline{\Delta S_{21}(\mathbf{k}'',\mathbf{k}) \Delta S_{21}(\mathbf{k}',\mathbf{k})} = \sigma_{21}(\mathbf{k}',\mathbf{k})\delta(\mathbf{k}'' - \mathbf{k}'). \quad (11)$$

Given that the ensemble average of the current density normal to junction in the metal $M'$ can be calculated using the formula $\overline{J_z} = (\hbar/m) Im\overline{(\psi_{M'}^* d\psi_{M'}/dz)}$, the transmission coefficient $D(E, \mathbf{k})$ appearing in Eq. 4 can now be found using Eqs. 8, 10 and 11, with the result that

$$D(E, \mathbf{k}) = |\overline{V}_{M',M}(\mathbf{k})|^2 + \int \sigma_{M',M}(\mathbf{k}',\mathbf{k}) d^2k', \quad (12)$$

where the integration is restricted to $|\mathbf{k}'| < K_M$. From Eq. 12 it is clear that the fraction of the incoming current in M with parallel wave vector $\mathbf{k}$ transmitted into states in $d^2k'$ is $\sigma_{M',M}(\mathbf{k}',\mathbf{k}) d^2k'$ whereas the fraction transmitted without change in the parallel component of momentum is $|\overline{V}_{M',M}(\mathbf{k})|^2$. In terms of quantities characterizing the two junction interfaces, one finds

$$|\overline{V}_{M',M}(\mathbf{k})|^2 = |\overline{V}_{M',I}(\mathbf{k})\overline{V}_{I,M}(\mathbf{k})|^2 \qquad (13)$$

and

$$\sigma_{M'M}(\mathbf{k}',\mathbf{k}) = \sigma_{M'I}(\mathbf{k}',\mathbf{k})|\overline{V}_{I,M}(\mathbf{k})|^2 e^{-2\kappa_{Ik}t}$$
$$+ |\overline{V}_{M',I}(\mathbf{k}')|^2 \sigma_{IM}(\mathbf{k}',\mathbf{k}) e^{-2\kappa_{Ik'}t}$$
$$+ \int d^2 k'' \sigma_{M'I}(\mathbf{k}',\mathbf{k}'') e^{-2\kappa_{Ik''}t} \sigma_{IM}(\mathbf{k}'',\mathbf{k}) \qquad (14)$$

The diffuse contribution to the tunneling current, Eq. 14, contains contributions in which the transmission is diffuse at one interface and specular at the other, as well as a contribution (the last term) which is diffuse at both interfaces.

*THE SMALL PERTURBATION METHOD* The small perturbation method works when the flat surface problem (i.e. the problem for $h_\alpha(\mathbf{r}) = 0$) is a good first approximation. The corrections are calculated by expanding in powers of $h_\alpha(\mathbf{r})$. The quantities $h_\alpha(\mathbf{r})$ appear in the calculations because expressions such as Eqs. 7 and 8 are evaluated at the interfaces $z = -t + h_1(\mathbf{r})$ and $z = h_2(\mathbf{r})$ when applying the boundary conditions. Thus $h_\alpha(\mathbf{r})$ appears in expressions such as $exp[-\kappa_{Ik} h_\alpha(\mathbf{r})]$ and $exp[iq_{Mk} h_\alpha(\mathbf{r})]$, and expansions in powers of $h_\alpha(\mathbf{r})$ will be expansions in powers of the parameters $[\kappa_{Ik} h_\alpha(\mathbf{r})]$ and $[q_{Mk} h_\alpha(\mathbf{r})]$. The quantities $q_{Mk}$ and $\kappa_{Ik}$ are of the order of magnitude of $2\pi/\lambda$ where the electron's wavelength $\lambda$ is expected to be comparable in magnitude to the lattice constant. The small perturbation approach will therefore be valid only when the root mean square fluctuations in $h_\alpha(\mathbf{r})$ are smaller than a lattice constant, i.e. for atomically flat interfaces. Since the results of the section on flat interfaces are already a good first approximation when the small perturbation method is applicable, no further results of this approximation will be given.

*THE TANGENT PLANE APPROXIMATION* This section evaluates the transmission coefficient $D(E,\mathbf{k})$ occurring in Eq. 4 for the tunneling current within the framework of the tangent plane approximation [9]. This approximation works best when the spatial scale of the roughness is larger than the electron wavelength, and is thus complementary to the small perturbation approach outlined in the previous section.

Consider first the general case of the transmission of a plane wave from medium 1 to medium 2 across the random interface $z = h(\mathbf{r})$, which is described in terms of Eqs. 7 and 8. The method begins with a mathematical formulation of Huygens' principle in which the wave function of the electron in the medium 2 is given in terms of its value and the value of its normal derivative on the interface $z = h(\mathbf{r})$, namely,

$$\psi_2(\mathbf{R}) = -\int \psi_2(\mathbf{R}') \frac{\partial G_0(\mathbf{R}' - \mathbf{R})}{\partial n'} dS'$$

$$+ \int \frac{\partial \psi_2(\mathbf{R}')}{\partial n'} G_0(\mathbf{R}' - \mathbf{R}) dS'. \qquad (15)$$

Here $S'$ is the surface $z = h(\mathbf{r})$, $\mathbf{R}'$ is on this surface, and $\partial/\partial n'$ is a normal derivative into medium 2. Also, the Green's function $G_0(\mathbf{R}' - \mathbf{R})$ satisfies the equation $(\nabla^2 + K_2^2)G_0(\mathbf{R}' - \mathbf{R}) = \delta(\mathbf{R}' - \mathbf{R})$ and can be represented as

$$G_0(\mathbf{R}) = -\frac{i}{8\pi^2} \int \frac{exp[-i\mathbf{k}\cdot\mathbf{r} + iq_{2k}|z|]}{q_{2k}} d^2k. \qquad (16)$$

The next step is to find the electron wave function, $\psi_2(\mathbf{R}')$, at points $z = h(\mathbf{r})$ on the interface. This is done in the tangent plane approximation by considering a given point on the interface, constructing a tangent plane there, and then considering the reflection and transmission of the incoming plane wave (which is taken to be the first term in Eq. 7) at this tangent plane. This gives $\psi_2(\mathbf{R}')$ in terms of the amplitude and phase of the incoming plane wave, and this result can be combined with Eqs. 8, 15 and 16 to yield

$$S_{21}(\mathbf{k}',\mathbf{k}) =$$
$$\int exp[-i(\mathbf{k}' - \mathbf{k})\cdot\mathbf{r} + i(q_{2k} - q_{1k})h(\mathbf{r})] \frac{d^2 r}{(2\pi)^2}. \qquad (17)$$

Here, a complicated function of the wave vectors of order unity, and analogous to the prefactor $g$ in Eq. 5, has been omitted.

The quantities $\overline{V}_{M',M}$ and $\sigma_{M',M}$ necessary for an evaluation of the transmission coefficient $D(E,\mathbf{k})$ (Eq. 12) can now be evaluated by combining Eqs. 9, 10, 11 and 17. In carrying out the necessary ensemble averages the function $h(\mathbf{r})$ is assumed to be Gaussian, and the theorem $\overline{exp(-h)} = exp(-\overline{h^2}/2)$ (valid for any Gaussian variable having $\overline{h} = 0$) is used. The results are

$$|\overline{V}_{M',M}(\mathbf{k})|^2 = F\beta_1\beta_2 P_{\Delta k_I}(\mathbf{k}) \qquad (18)$$

and

$$\sigma_{M',M}(\mathbf{k}',\mathbf{k}) = F\beta_1 P_{\Delta k_2}(\mathbf{k}' - \mathbf{k})P_{\Delta k_I}(\mathbf{k})$$
$$+ F\beta_2 P_{\Delta k_I}(\mathbf{k}')P_{\Delta k_1}(\mathbf{k}' - \mathbf{k})$$
$$+ F\int d^2 k'' P_{\Delta k_2}(\mathbf{k}' - \mathbf{k}'')P_{\Delta k_I}(\mathbf{k}'')P_{\Delta k_1}(\mathbf{k}'' - \mathbf{k}). \qquad (19)$$

where

$$F = 2\pi(\Delta k_I)^2 e^{-2\kappa_I t} e^{2\kappa_I^2(\overline{h_1^2} + \overline{h_2^2})}, \qquad (20)$$
$$\beta_\alpha = e^{-(\kappa_I^2 + K_M^2)\overline{h_\alpha^2}}, \qquad (21)$$
$$(\Delta k_\alpha)^2 = (\kappa_I^2 + K_M^2)\overline{s_\alpha^2}. \qquad (22)$$

The $P_{\Delta k_\alpha}$'s are the normalized Gaussian functions defined by Eq. 6 with widths $\Delta k_\alpha$ given by Eq. 22 where $\overline{s_\alpha^2} = \overline{(\partial h_\alpha/\partial x)^2} = \overline{(\partial h_\alpha/\partial y)^2}$ is the mean square slope of the roughness.

*SUMMARY AND CONCLUSIONS* The approach to tunneling theory introduced above allows a calculation of the consequences of rough tunnel junction interfaces on the tunneling current and on its directionality. The general formula for the tunneling current is given by Eqs. 4, 12, 13, and 14. Eq. 12 shows the separation of the current into specular and diffuse parts, and the scattering cross section $\sigma_{M'M}$ gives the directional dependence of the diffuse part. Eqs. 13, and 14 reduce the problem to the determination of the transmission properties of the individual junction interfaces. These expressions give a formally exact theory of tunneling for thick rough tunnel junctions, and can be evaluated using any appropriate approximation scheme.

The small perturbation method treats the roughness as a perturbation of a flat interface model, and shows that flat interface models represent a good first approximation when the amplitude of the roughness fluctuations is less than the electron wavelength (which normally requires atomically flat interfaces).

The results obtained in the tangent plane approximation show that for rough tunnel junction interfaces (i.e. surface height fluctuations significantly greater than the electron wavelength) the transmitted current is nearly totally diffuse. To see this recall that the functions $P$ occurring in Eqs. 18 and 19 are normalized Gaussians. Thus the relative weights of the different contributions in Eqs. 18 and 19 to the tunneling current are determined by the factors $\beta_\alpha$. This means that for root mean square fluctuations in the height functions $h(\mathbf{r})$ much greater than an electron wavelength, i.e. such that the factors $\beta_\alpha$ are small, the purely diffuse contribution, namely that last term in Eq. 19, dominates.

Now examine the directionality in the rough junction case where the tunneling current is dominated by the last term in Eq. 19. For sufficiently thick tunnel junctions, the factor $P_{\Delta k_I}(\mathbf{k}'') = \delta(\mathbf{k}'')$ and the integration over $\mathbf{k}''$ is easily carried out. The incoming and outgoing electrons contributing to the tunneling current thus have their parallel moment components within $\Delta k_1$ and $\Delta k_2$ (see Eq. 22) of zero, respectively. For root mean square (rms) roughness slopes $s_\alpha$ which are of order unity or not too much smaller, there is no significant directionality of the tunneling. On the other hand for rms roughness slopes much less than unity, the smaller the roughness slope, the closer to the forward direction is the momentum of electrons contributing to the tunneling current, both for incoming and outgoing electrons.

It is of interest to examine the physical reasons for the dominance of the diffuse component of the tunneling current in the case of rough junctions. As for flat junctions, the tunneling current is reduced by the factor $exp(-2\kappa_I t)$ depending exponentially on the average thickness $t$ of the junction, (see Eq. 20). This effect is reduced by the interface height fluctuations, which give regions where the potential barrier has a smaller than average thickness. Hence the factor $exp(2\kappa_I^2(\overline{h_1^2} + \overline{h_2^2}))$ in Eq. 20. The reduction of the attenuation due to barrier thickness fluctuations is not as great for the specular component of the transmission, as indicated by the factor $exp(-\kappa_I^2 \overline{h_\alpha^2})$ in $\beta_\alpha$. The other factor contributing to $\beta_\alpha$, $exp(-K_M^2 \overline{h_\alpha^2})$, gives the reduction in the specularly transmitted component of transmission due to destructive interference of waves with the different phase lags due to having traveled different distances in the insulator. These two factors combine to make the specular component of the tunneling negligible relative to the diffuse component for sufficiently rough interfaces. Clearly, tunneling theory must account for the roughness of tunnel junctions in order to correctly describe the dominant diffuse contribution to the tunneling current.

This article has given a formally exact expression for the tunneling current valid for thick, rough tunnel junctions, has shown that for rough tunnel junctions the diffuse component of the tunneling current dominates the specular component, and has also shown that even when the tunneling current is entirely diffuse, a tunneling cone effect can exist if the root mean square roughness slope is sufficiently small.

*ACKNOWLEDGEMENTS* I wish to thank M. Aprili, J.P. Carbotte, and J.R. Kirtley for stimulating discussions, and the Natural Sciences and Engineering Research Council of Canada for support.